\documentclass[11pt,a4paper]{article}
\begin{document}
\title{Can an observer really catch up with light? \footnote{E-mail of Tian:tgh-2000@263.net.}}
\author{Guihua Tian\\Department
of Physics, Beijing Normal University, Beijing100875, China.\\
School of Science, Beijing University \\
of Posts And Telecommunications. Beijing100876, China.\\
Zhao Zheng,\\ Department of Physics, Beijing Normal University,
Beijing100875, China.}
\date{\today}
\maketitle
\begin{abstract}
Given a null geodesic $\gamma _0(\lambda)$ with a point $r$ in
$(p,q)$ conjugate to $p$ along $\gamma _0(\lambda)$ , there will
be a variation of $\gamma _0(\lambda)$ which will give a time-like
curve from $p$ to $q$. This is a well-known theory proved in the
famous book\cite{2}. In the paper we prove that the time-like
curves coming from the above-mentioned variation have a proper
acceleration which approaches infinity as the time-like curve
approaches the null geodesic. This means no observer can be
infinitesimally near the light and begin at the same point with
the light and finally catch the light. Only separated from the
light path finitely, does the observer can begin at the same point
with the light and finally catch the light.
\end{abstract}

It is well-known that an observer in "hyperbolic" motion in
Minkowski space-time has a constant proper acceleration (the
magnitude of the 4-acceleration). The equations of the world line
of one of such observers can be expressed concisely as $x=x_0$,
$y=y_0$, $z=z_0$, (and the proper acceleration $A=x_0^{-1}$.)
where {$t,\ x,\ y,\ z$} are the Rindler coordinates, and the line
element in these coordinates of the Minkowski metric reads
\begin{eqnarray}
ds^2=-x^2dt^2+dx^2+dy^2+dz^2.\\
 (-\infty<t<+\infty,\ x>0,\
-\infty<y<+\infty ,\ -\infty<z<+\infty .)\nonumber
\end{eqnarray}

Consider a one-parameter family of hyperbolic observers ($x_0$,
being the parameter) with the same $y_0$ and $z_0$, then the
proper acceleration, $A=x_0^{-1}$, approaches infinity as $x_0$
approaches zero. As a limit case, the curve defined by $x=0$,
$y=y_0$, $z=z_0$ is a null geodesic. Unlike time-like curves, the
concept of 4-acceleration of a null geodesic (or even a null
curve), to our knowledge, has not been defined. The fact that $A
\rightarrow  \infty $ as $x_0 \rightarrow 0$, however, suggests
that it seems not unreasonable to define the proper acceleration
of a null geodesic (or physically, a photon ) in Minkowski
space-time to be infinity. This is indeed what Rindler suggested
in his book \cite{1}. The main purpose of the previous
paper\cite{4} is to generalize this result to curved space-times,
namely, to argue that it is not unreasonable to define the proper
acceleration of a null geodesic which is future-complete in curved
space-time to be infinity. The present paper continues the main
point of the paper \cite{4}, and primely concerns the proper
acceleration of time-like curve coming from the variation of null
geodesic with two end points fixed on the null geodesic, and gives
the conclusion that the proper acceleration of this type of
time-like curve does approaches infinity as the time-like curve
approaches the null geodesic. Because the two end points fixed on
the null geodesic, the existence of the time-like curves from
variation of $\gamma _0$ is in question. There are two theorems
that concern the existence of the time-like curves:

theorem1 \cite{2},\cite{3}. Let $\gamma _0(\lambda)$ be a smooth
causal curve and let $p,\ q\in$ $\gamma _0(\lambda)$. Then there
does not exist a smooth one-parameter family of causal curves
$\sigma \left( u,\lambda \right)$ connecting $p$ and $q$ with
$\sigma \left( 0,\lambda \right)=\gamma _0(\lambda)$ and $\gamma
_u(\lambda)$ time-like for all $u >0$ (i.e., $\gamma _0(\lambda)$
cannot be smoothly deformed to a time-like curve) if and only if
$\gamma _0(\lambda)$ is null geodesic with no point conjugate to
$p$ along $\gamma _0(\lambda)$ between $p$ and $q$.

theorem2\cite{2},\cite{3}.If there is a point $r$ in $(p,q)$
conjugate to $p$ along $\gamma _0(\lambda)$, then there will be a
variation of $\gamma _0(\lambda)$ which will give a time-like
curve from $p$ to $q$.

We therefore suppose that $\gamma _0(\lambda)$ is  a null geodesic
with a point $r$ in $(p,q)$ conjugate to $p$ along $\gamma
_0(\lambda)$ and the existence of the time-like curves connecting
$p$,$q$ obtained from variation is ensured. Precisely ,we have the
following definition of the variation of $\gamma _0(\lambda)$:

Let ($M,g_{ab}$) be a 4-dimensional curved space-time and $\gamma
_0 $: ($0,\lambda_{q}$)$\rightarrow M$ be a null geodesic, which
will later be denoted by $\gamma _0(\lambda)$ with $\lambda$ its
affine parameter, and with $p,\ q\in$ $\gamma _0(\lambda)$. We
define a variation of $\gamma _0 $ to be a $C^{1-}$  map \cite{2}
$\sigma$ :$ \left[0,\varepsilon \right) \times
\left[0,\lambda_{q}\right] \rightarrow M $
such that \\

(1)$\sigma \left( 0,\lambda \right) =\gamma _0\left( \lambda
\right) $,

(2)$\sigma \left( u,0 \right)=p$, $\sigma \left( u,\lambda_{q}
\right)=q$

(3)there is a subdivision $0=\lambda_1<\lambda_2<
...<\lambda_n=\lambda_q$ of $[0,\lambda_q]$ such that $\sigma $ is
$C^3$ on each $\left[0,\varepsilon \right) \times \left[\lambda _i
,\lambda_{i+1}\right]$.

(4)for each constant $u\in \left[0,\varepsilon \right) $ and $u \neq 0$, $%
\sigma \left( u,\lambda \right) $ is a time-like curve and is
represented by $\gamma _u\left( \lambda \right)$.

Denote by $\left( \frac \partial {\partial \lambda }\right)
_u^a\equiv v_u^a$ the tangent vector to the curve
$\gamma_u(\lambda )$, then $\left( \frac \partial {\partial
\lambda }\right) _0^a\equiv v_0^a$ satisfies the null geodesic
equation:
\begin{equation}
\left( \frac \partial {\partial \lambda }\right) _0^b\nabla
_b\left( \frac
\partial {\partial \lambda }\right) _0^a=0 ,\label{null geod1}
\end{equation}
where $\nabla _a$ is the unique derivative operator associated
with $g_{ab}$, i.e., $\nabla _a g_{bc}=0$. We rewrite
eq.(\ref{null geod1})as
\begin{equation}
v_0^b\nabla _bv_0^a=0 \label{null geod2}
\end{equation}

 Let $\left( \frac \partial {\partial u}\right) ^a$ be the tangent vector to the
curve $\sigma (u,\lambda)$ with $\lambda =const$, and define the
variation vector field $Z^a$ on $\gamma_0(\lambda)$ by
\begin{equation}
Z^a=\left( \frac \partial {\partial u}\right) ^a|_{u=0},
\end{equation}
then it is not difficult to see that the Lie derivative of $\left(
\frac
\partial {\partial u}\right) ^a$ with respect to  $\left( \frac
\partial {\partial \lambda}\right) ^a$ vanishes
\cite{2},i.e.,
\begin{equation}
L _{\frac \partial {\partial \lambda}}\left( \frac
\partial {\partial u}\right) ^a=0 \label{li}.
\end{equation}
that is
\begin{equation}
v_u^b\nabla _b\left(\frac{\partial}{\partial u} \right)^a=\left(
\frac
\partial {\partial u }\right) ^b\nabla _bv_u^a \label{li2},
\end{equation}
evaluation of equation (\ref{li2}) on $\gamma_0(\lambda )$ yields
\begin{equation}
v_0^b\nabla _bZ^a=Z ^b\nabla _bv_0^a\label{li3},
\end{equation}
If we denote $g_{ab}v_u^av_u^b$ by $-\alpha _u^2$, that is
\begin{equation}
-\alpha _u^2=g_{ab}v_u^av_u^b\label{2derivative1},
\end{equation}
and decompose $-\alpha _u^2$ into Taylor series
\begin{equation}
-\alpha _u ^2=g_{ab}v_u^av_u^b=-\alpha _0^2+\beta_1u+\frac 12\beta
_2 u^2+0(u^3),
\end{equation}
where
\begin{equation}
\alpha _0^2=g_{ab}v_0^av_0^b=0.
\end{equation}
In order to get $\gamma_u$to be time-like, it is easy to see
$\beta_1=\frac{\partial (-\alpha ^2)}{\partial u}|_{u=0}\leq 0$
and it can be prove $\beta_1=0$ (see detail in reference
\cite{2}). The following is the reason: with eq(\ref{li2})
\begin{eqnarray}
\frac{\partial (-\alpha _u^2)}{\partial
u}&=&(\frac{\partial}{\partial u})^c\nabla
_c\left[g_{ab}v_u^av_u^b\right]=2g_{ab}v_u^a(\frac{\partial}{\partial
u})^c\nabla _cv_u^b=2g_{ab}v_u^av_u^c\nabla
_c(\frac{\partial}{\partial u})^b \nonumber \\
&=&2v_u^c\nabla _c\left[g_{ab}v_u^a(\frac{\partial}{\partial
u})^b\right]-2g_{ab}(\frac{\partial}{\partial u})^bv_u^c\nabla
_cv_u^a\nonumber\\ &=&2\frac{\partial }{\partial \lambda
}\left[g_{ab}v_u^a(\frac{\partial}{\partial
u})^b\right]-2g_{ab}(\frac{\partial}{\partial u})^bv_u^c\nabla
_cv_u^a,\label{1derivative1}
\end{eqnarray}
therefore, with eq.(\ref{null geod2}), one gets
\begin{equation}
\beta_1=\frac{\partial (-\alpha _u^2)}{\partial
u}|_{u=0}=2\frac{\partial }{\partial \lambda
}\left[g_{ab}v_0^aZ^b\right]=2\frac{dh}{d\lambda}\label{1derivative2}
\end{equation}
where $h(\lambda)=g_{ab}v_0^aZ^b$ is continuous on $(0,\lambda_q)$
thanks to the continuity of of $Z^a$ . From the property of the
variation map $\sigma(u,\lambda)$, that is , $\sigma \left( u,0
\right)=p$, $\sigma \left( u,\lambda_{q} \right)=q$, we get
\begin{equation}
Z^a(0)=Z^a(\lambda_q)=0\label{zz}.
\end{equation}
This in turn induces $h(0)=h(\lambda_q)=0$, but
$h(0)=h(\lambda_q)=0$ is impossible if $\beta_1$ is less than
zero. So, $\beta_1$ must be zero to make $\gamma _u$ time-like.

$\beta_1=0$ induces $h=g_{ab}v_0^aZ^b=0$, that is, the variation
vector $Z^a$ is orthogonal to the tangent vector $v_0^a$ of the
null geodesic $\gamma_0$.

Therefore, one gets
\begin{equation}
-\alpha _u ^2=g_{ab}v_u^av_u^b=\frac 12\beta _2
u^2+0(u^3)\label{xianchang},
\end{equation}

$\beta_2=\frac12\frac{\partial ^2(-\alpha _u^2)}{\partial
u^2}|_{u=0}$ is followed from eq.(\ref{1derivative1})(see detail
in reference \cite{2})

\begin{eqnarray}
&& \frac12\frac{\partial ^2 (-\alpha _u^2)}{\partial
u^2}=\frac{\partial ^2 }{\partial \lambda \partial
u}\left[g_{ab}v_u^a(\frac{\partial}{\partial
u})^b\right]-(\frac{\partial}{\partial u})^d\nabla _d\left[
g_{ab}(\frac{\partial}{\partial u})^bv_u^c\nabla _cv_u^a\right]\nonumber\\
&=&\frac{\partial ^2 }{\partial \lambda \partial
u}\left[g_{ab}v_u^a(\frac{\partial}{\partial
u})^b\right]-\left[g_{ab}v_u^c\nabla
_cv_u^a(\frac{\partial}{\partial u})^d\nabla
_d(\frac{\partial}{\partial u})^b\right]-\left[
(\frac{\partial}{\partial u})^a(\frac{\partial}{\partial
u})^d\nabla _d\left(v_u^c\nabla _cv_{ua}\right) \right]
\end{eqnarray}
the term $(\frac{\partial}{\partial u})^d\nabla _d(v_u^c\nabla
_cv_{ua})$ in the third part of above equation is simplified:
\begin{eqnarray*}
&&(\frac{\partial}{\partial u})^d\nabla _d\left(v_u^c\nabla
_cv_{ua}\right)=\left((\frac{\partial}{\partial u})^d\nabla
_dv_u^c\right)\nabla _cv_{ua}+v_u^c(\frac{\partial}{\partial
u})^d\nabla _d\nabla _cv_{ua}\\
&& =\left((\frac{\partial}{\partial u})^d\nabla
_dv_u^c\right)\nabla _cv_{ua}+v_u^c(\frac{\partial}{\partial
u})^d\nabla _c\nabla
_dv_{ua}+R_{dcae}v_u^cv_u^e(\frac{\partial}{\partial u})^d \\
&&=\left((\frac{\partial}{\partial u})^d\nabla
_dv_u^c\right)\nabla _cv_{ua}+v_u^c\nabla
_c\left[(\frac{\partial}{\partial u})^d\nabla _dv_{ua}\right]\\
&&-\left(v_u^c\nabla _c(\frac{\partial}{\partial
u})^d\right)\nabla
_dv_{ua}+R_{dcae}v_u^cv_u^e(\frac{\partial}{\partial u})^d \\
&&=v_u^d\nabla _d\left(v_u^c\nabla _c(\frac{\partial}{\partial
u})_a\right)+R_{dcae}v_u^cv_u^e(\frac{\partial}{\partial u})^d,
\end{eqnarray*}
that is
\begin{equation}
(\frac{\partial}{\partial u})^d\nabla
_d\tilde{A}_{ua}=(\frac{\partial}{\partial u})^d\nabla
_d\left(v_u^c\nabla _cv_{ua}\right)=v_u^d\nabla
_d\left(v_u^c\nabla _c(\frac{\partial}{\partial
u})_a\right)+R_{dcae}v_u^cv_u^e(\frac{\partial}{\partial
u})^d.\label{hatAderivative}
\end{equation}
where  the relation $\nabla _c\nabla _dv_{ua}-\nabla _d\nabla
_cv_{ua}=-R_{dcae}v_u^e$ in second step and eq.(\ref{li2})in the
fourth step have been used, and $\tilde{A}_{ua}$ is defined by
$\tilde{A}_{ua}=v_u^c\nabla _cv_{ua}$ and different from the
proper acceleration of the time-like curve $\gamma_u(\lambda)$
(see following for the definition of the proper acceleration of
$\gamma_u(\lambda)$). Therefore, one gets
\begin{eqnarray}
&&\frac12\frac{\partial ^2 (-\alpha _u^2)}{\partial
u^2}=\frac{\partial ^2 }{\partial \lambda \partial
u}\left[g_{ab}v_u^a(\frac{\partial}{\partial
u})^b\right]-\left[g_{ab}\left(v_u^c\nabla
_cv_u^a\right)(\frac{\partial}{\partial u})^d\nabla
_d(\frac{\partial}{\partial u})^b\right]\nonumber\\
&&-\left[(\frac{\partial}{\partial u})^av_u^d\nabla
_d\left(v_u^c\nabla _c(\frac{\partial}{\partial
u})_a\right)+R_{dcae}v_u^cv_u^e(\frac{\partial}{\partial
u})^a(\frac{\partial}{\partial u})^d\right],
\end{eqnarray}
so,
\begin{equation}
\beta _2=\left[\frac12\frac{\partial ^2 (-\alpha _u^2)}{\partial
u^2}\right]_{u=0}=\left[\frac{\partial ^2 }{\partial \lambda
\partial u}\left(g_{ab}v_u^a(\frac{\partial}{\partial
u})^b\right)\right]_{u=0}-\left[Z^av_0^d\nabla _d\left(v_0^c\nabla
_cZ_a\right)+R_{dcae}v_0^cv_0^eZ^dZ^a\right]\label{beta2},
\end{equation}
where eq. (\ref{null geod2}) has been used. for simplicity, one
has
\begin{eqnarray}
\tilde{\beta} _2&=& \left[\frac{\partial ^2}{\partial \lambda
\partial u}(g_{ab}v_u^a(\frac{\partial}{\partial
u})^b)\right]_{u=0}= \frac{\partial }{\partial
\lambda}\left[(\frac{\partial}{\partial u})^c\nabla _c
(g_{ab}v_u^a(\frac{\partial}{\partial
u})^b)\right]_{u=0}\nonumber\\
&=&\frac{\partial }{\partial
\lambda}\left[g_{ab}(\frac{\partial}{\partial
u})^b(\frac{\partial}{\partial u})^c\nabla _c v_u^a\right]_{u=0}+
\frac{\partial }{\partial
\lambda}\left[g_{ab}v_u^a(\frac{\partial}{\partial u})^c\nabla _c
(\frac{\partial}{\partial u})^b\right]_{u=0}\nonumber\\
&=&\frac{\partial }{\partial
\lambda}\left[g_{ab}(\frac{\partial}{\partial u})^bv_u^c\nabla
_c(\frac{\partial}{\partial
u})^a\right]_{u=0}=\frac12\frac{\partial ^2}{\partial \lambda ^2
}\left[g_{ab}(\frac{\partial}{\partial
u})^b(\frac{\partial}{\partial u})^a\right]_{u=0}\nonumber\\
&=&\frac 12 \frac{d ^2}{d\lambda
^2}\left(g_{ab}Z^aZ^b\right)\label{foot1},
\end{eqnarray}
where the relation $\left[(\frac{\partial}{\partial u})^c\nabla _c
(\frac{\partial}{\partial u})^b\right]_{u=0}=0$ has been used,
which comes from the relation of the variation map
$\sigma(u,\lambda)$ and the variation vector $Z^a$:
$\sigma(u,\lambda)=exp_r(uZ^a),\ r=\gamma_0(\lambda)$ with
$u=const$ (see page 107 of the reference \cite{2}).
\begin{equation}
\beta _2=\frac 12 \frac{d ^2}{d\lambda
^2}\left(g_{ab}Z^aZ^b\right)-\left[Z^av_0^d\nabla
_d\left(v_0^c\nabla
_cZ_a\right)+R_{dcae}v_0^cv_0^eZ^dZ^a\right]\label{beta21}.
\end{equation}

The parameter $\lambda$ of the timelike curve, $\gamma _u\left(
\lambda \right) $, defined above is not, in general, the proper
time of the curve. If one re-parameterizes the curve $\gamma
_u\left( \lambda \right) $ by its proper time $\tau$, i.e., the
parameter satisfying
\begin{equation}
g_{ab}\left( \frac \partial {\partial \tau }\right) _u^a\left(
\frac
\partial {\partial \tau }\right) _u^b=-1,
\end{equation}
then \begin{equation}
\left( \frac \partial {\partial \tau }\right) ^a=\left( \frac{d\lambda }{%
d\tau }\right) \left( \frac \partial {\partial \lambda }\right)
_u^a=\left( \frac{d\lambda }{%
d\tau }\right)v_u^a.
\end{equation}
With eq.(\ref{2derivative1}), one has
\begin{equation} \left(
\frac{d\lambda }{d\tau }\right) ^2=\frac 1{\alpha_u ^2}.
\end{equation}

The 4-acceleration of the time-like curve $\gamma _u$ which is
defined as
\begin{equation}
A^a=\left( \frac \partial {\partial \tau }\right) ^b\nabla
_b\left( \frac
\partial {\partial \tau }\right) ^a=\frac{d\lambda }{d\tau }
v_u^b \nabla _b\left(\frac{d\lambda }{d\tau }v_u^a\right)=\left(
\frac 1{\alpha _u^2}\right)\widetilde{A}_u^a-\frac 1{2\alpha _u
^4}v_u^av_u^b\nabla _b\alpha _u^2, \label{ac.},
\end{equation}
or, one writes the above equation as
\begin{equation}
\widetilde{A}_u^a= \alpha _u^2A^a+\frac 1{2\alpha _u
^2}v_u^av_u^b\nabla _b\alpha _u^2, \label{acjia.}.
\end{equation}
with eq.(\ref{acjia.}), then
\begin{equation}
(\frac{\partial}{\partial u})^a(\frac{\partial}{\partial
u})^d\nabla _d (\alpha _u^2A_a)=(\frac{\partial}{\partial
u})^a(\frac{\partial}{\partial u})^d\nabla
_d\tilde{A}_{ua}-(\frac{\partial}{\partial
u})^a(\frac{\partial}{\partial u})^d\nabla _d\left(\frac 1{2\alpha
_u ^2}v_{ua}v_u^b\nabla _b\alpha _u^2\right),
\end{equation}
first, the calculation of $b_1\equiv(\frac{\partial}{\partial
u})^a(\frac{\partial}{\partial u})^d\nabla _d\left(\frac 1{2\alpha
_u ^2}v_{ua}v_u^b\nabla _b\alpha _u^2\right)$ is
\begin{eqnarray}
b_1&=&(\frac{\partial}{\partial
u})^av_{ua}(\frac{\partial}{\partial u})^d\nabla _d\left(\frac
1{2\alpha _u ^2}v_u^b\nabla _b\alpha _u^2\right)+\frac 1{2\alpha
_u ^2}(v_u^b\nabla _b\alpha _u^2)(\frac{\partial}{\partial
u})^a(\frac{\partial}{\partial
u})^d\nabla _dv_{ua}\nonumber\\
&=&(\frac{\partial}{\partial u})^av_{ua}(\frac{\partial}{\partial
u})^d\nabla _d\left(\frac 1{2\alpha _u ^2}v_u^b\nabla _b\alpha
_u^2\right)+\frac 1{2\alpha _u ^2}(v_u^b\nabla _b\alpha
_u^2)(\frac{\partial}{\partial u})^av_u^d\nabla
_d(\frac{\partial}{\partial u})_{a}\nonumber\\
&=&(\frac{\partial}{\partial u})^av_{ua}(\frac{\partial}{\partial
u})^d\nabla _d\left(\frac 1{2\alpha _u ^2}v_u^b\nabla _b\alpha
_u^2\right)+\frac 1{4\alpha _u ^2}(v_u^b\nabla _b\alpha
_u^2)v_u^d\nabla _d[(\frac{\partial}{\partial
u})_{a}(\frac{\partial}{\partial u})^a].
\end{eqnarray}
with eq. (\ref{zz}), therefore,
\begin{equation}
\lim_{u\rightarrow 0}b_1=\lim_{u\rightarrow 0}(\frac 1{4\alpha _u
^2}v_u^b\nabla _b\alpha _u^2)v_0^d\nabla
_d(Z^aZ_a)=\lim_{u\rightarrow 0}(\frac 1{4\alpha _u ^2}v_u^b\nabla
_b\alpha _u^2)\frac{d}{d\lambda }(Z^aZ_a)\label{b1}
\end{equation}
second, with eqs.(\ref{hatAderivative}),(\ref{beta2}),
(\ref{beta21}), calculate $b_2\equiv (\frac{\partial}{\partial
u})^a(\frac{\partial}{\partial u})^d\nabla _d\tilde{A}_{ua} $
\begin{eqnarray}
b_2&=&(\frac{\partial}{\partial u})^a[v_u^d\nabla
_d\left(v_u^c\nabla _c(\frac{\partial}{\partial
u})_a\right)+R_{dcae}v_u^cv_u^e(\frac{\partial}{\partial u})^d]\nonumber\\
\lim_{u\rightarrow 0}b_2&=& \left[Z^av_0^d\nabla
_d\left(v_0^c\nabla
_cZ_a\right)+R_{dcae}v_0^cv_0^eZ^dZ^a\right]\nonumber\\
&=&\frac 12 \frac{d ^2}{d\lambda
^2}\left(g_{ab}Z^aZ^b\right)-\beta _2\label{b2}.
\end{eqnarray}
thirdly, $b\equiv (\frac{\partial}{\partial
u})^a(\frac{\partial}{\partial u})^d\nabla _d (\alpha _u^2A_a)$ is
\begin{eqnarray}
b&=&(\frac{\partial}{\partial u})^d\nabla _d (\alpha
_u^2A_a(\frac{\partial}{\partial u})^a)-(\alpha
_u^2A_a)(\frac{\partial}{\partial u})^d\nabla
_d(\frac{\partial}{\partial u})^a\nonumber\\
\lim_{u\rightarrow 0}b&=&\lim_{u\rightarrow
0}(\frac{\partial}{\partial u})^d\nabla _d (\alpha
_u^2A_a(\frac{\partial}{\partial u})^a)\label{b},
\end{eqnarray}
where one uses the relation $\lim_{u\rightarrow
0}[(\frac{\partial}{\partial u})^d\nabla
_d(\frac{\partial}{\partial u})^a]=0$. The following shows that
$\lim_{u\rightarrow 0}b=\lim_{u\rightarrow
0}b_2-\lim_{u\rightarrow 0}b_1\neq 0$. With eqs.(\ref{xianchang}),
(\ref{b1}),(\ref{b2}), then
\begin{eqnarray}
\lim_{u\rightarrow 0}b&=&\lim_{u\rightarrow
0}b_2-\lim_{u\rightarrow 0}b_1=\frac 12 \frac{d ^2}{d\lambda
^2}\left(g_{ab}Z^aZ^b\right)-\beta _2-\lim_{u\rightarrow 0}(\frac
1{4\alpha _u ^2}v_u^b\nabla _b\alpha _u^2)\frac{d}{d\lambda
}(Z^aZ_a)\\
&=&\frac 12 \frac{d ^2}{d\lambda
^2}\left(g_{ab}Z^aZ^b\right)-\frac 1{4\beta _2
}\frac{d\beta_2}{d\lambda }\frac{d}{d\lambda }(Z^aZ_a)-\beta
_2\\
&=&\frac12\frac{(-\beta_2)}{(-\beta_2)^{\frac12}}\frac{d}{d\lambda}
\left[\frac{\frac{d}{d\lambda
}(Z^aZ_a)}{(-\beta_2)^{\frac12}}\right]-\beta_2\label{guanxi1}.
\end{eqnarray}
If $\lim_{u\rightarrow 0}b=0$, from eq.(\ref{guanxi1}), one
obtains $\frac12\frac{d}{d\lambda}\left[\frac{\frac{d}{d\lambda
}(Z^aZ_a)}{(-\beta_2)^{\frac12}}\right]= -(-\beta_2)^{\frac12}$.
Because $Z^a$ is a continuous, piecewise $C^2$  vector field along
$\gamma_0(\lambda )$ vanishing at end points p and q, $\frac
{d(Z^aZ_a)}{d\lambda}=0$ must be satisfied at end points p and q,
and in the first interval $[0,\lambda_2]$ ,$Z^a$ is $C^2$, then
$\forall \lambda \in [0,\lambda_2]$,
\begin{eqnarray}
&&\int_{0}^{\lambda}-2(-\beta_2)^{\frac12}d\lambda=\left[\frac{\frac{d}{d\lambda
}(Z^aZ_a)}{(-\beta_2)^{\frac12}}\right]_{\lambda},\\
&&\left[\frac{d(Z^aZ_a)}{d\lambda
}\right]_{\lambda}=-2(-\beta_2)^{\frac12}(\lambda)\int_{0}^{\lambda}(-\beta_2)^{\frac12}d\lambda
\end{eqnarray}
because $\gamma_u(\lambda)$ is time-like curve,
$(-\beta_2)^{\frac12}>0$ is satisfied everywhere, this in turn
ensures $\frac{d}{d\lambda }(Z^aZ_a)<0$ everywhere in the interval
$[0,\lambda_2]$ . This is impossible, as
$(Z^aZ_a)\mid_{\lambda=0}=0$, in the neighborhood of the initial
point p, $Z^a$ varies from zero to $Z^a\neq 0$, and $Z^a$ is
space-like, so, $(Z^aZ_a)$ increases with $\lambda $  in the
neighborhood of the initial point p , that is , $\frac{d}{d\lambda
}(Z^aZ_a)$ must be larger than zero in the neighborhood of the
initial point p. Therefore, $\lim_{u\rightarrow
0}b=\lim_{u\rightarrow 0}b_2-\lim_{u\rightarrow 0}b_1\neq 0$ is
guaranteed. From eq.(\ref{b}), one gets $\lim_{u\rightarrow
0}(\frac{\partial}{\partial u})^d\nabla _d (\alpha
_u^2A_a(\frac{\partial}{\partial u})^a)$ is not equal to zero, as
$\alpha _u^2=\frac12\beta_2 u^2$ approaches zero when
$u\rightarrow 0$, these insures that $A^aZ_a$ approaches infinity
as $u\rightarrow 0$, which induce $A^a\rightarrow \infty$ due to
the finiteness of the variation vector $Z^a$.

In conclusion, When a null geodesic $\gamma _0$ connecting $p$,
$q$ with a point $r\in (p,q)$ conjugate to $p$, then the proper
acceleration of  the time-like curve from $p$ to $q$ produced from
the variation of the $\gamma _0$ approaches infinity as
$u\rightarrow 0$, this means no observer can be infinitesimally
near the light and begin at the same point with the light and
finally catch the light. Only separated from the light path
finitely, does the observer can begin at the same point with the
light and finally catch the light.

\section{\label{section VI}Acknowledgments}

 We are supported by the National Science Foundation of China under Grant No.
10073002.

\end{document}